\documentclass[a4paper,11pt]{article}
\usepackage{jheppub}
\usepackage{amsmath}
\usepackage{multicol}
\usepackage{multirow}
\usepackage{subfigure}
\usepackage{slashed} 
\newcommand{\met}{\ensuremath{\slashed{E}_T}}
\newcommand{\wwa}{{\sc $W^+W^-\gamma$ }}
\newcommand{\wza}{{\sc $W^{\pm}Z\gamma$ }}
\newcommand{\wwaa}{{\sc $WW\gamma\gamma$ }}
\newcommand{\wwza}{{\sc $WWZ\gamma$ }}

\newcommand{\fbinv} {\mbox{\ensuremath{\,\text{fb}^\text{$-$1}}}}

\newcommand{\pythia}{{\sc Pythia}}
\newcommand{\delphes}{{\sc Delphes}}

\newcommand{\mgme}{{\sc MadGraph/MadEvent}}

\newcommand{\madgraph}{{\sc MadGraph}}
\newcommand{\tauola}{{\sc TAUOLA}}
\newcommand{\madevent}{{\sc MadEvent}}

\title{The CERN LHC Sensitivity on measuring \wza Production and Anomalous \wwza Coupling}

\author{Ke Ye,}
\author{Daneng Yang,}
\author{Qiang Li}

\affiliation{Department of Physics and State Key Laboratory of Nuclear Physics and Technology, \\
Peking University, Beijing, 100871, China}

\emailAdd{kevinye@pku.edu.cn, pmydn@pku.edu.cn, qliphy0@pku.edu.cn}

\abstract{In this paper we present for the first time a detailed Monte Carlo study of measuring \wza production with pure leptonic decays and probing anomalous quartic gauge-boson \wwza couplings at the $\sqrt{s}=14$ TeV LHC, with parton shower and detector simulation effects taken into account. We find that with an integrated luminosity of 100 $\fbinv$ and proper selection cuts, the Standard Model \wza signal significance can be improved to as much as $3\sigma$. After reviewing previous parametrization on anomalous \wwza couplings (see e.g. $a_n/\Lambda^2$ or  $k_2^m/\Lambda^2$ as shown in Ref.~\cite{Belanger:2000}),
we propose a more general parametrization scheme with 4 free inputs leading only to genuine \wwza aQGC couplings. Finally, our numerical results show that one can reach constraints at 95\% confidence level of $-5.7 \times 10^{-5}$ GeV$^{-2}$ $< k_2^m/\Lambda^2 <$ $5.5 \times 10^{-5}$ GeV$^{-2}$ and $-2.2 \times 10^{-5}$ GeV$^{-2}$ $< a_n/\Lambda^2 <$ $2.4 \times 10^{-5}$ GeV$^{-2}$, which are more stringent than LEP's results by three orders of magnitude.}

\date{\Date}


\begin{document}
\maketitle
\flushbottom
%---------------------------------------------------------------------

\section{Introduction}
\label{intro}
The Standard Model (SM) has so far undergone considerable experimental tests and proved to be quite successful, especially after the recent discovery of the 125-126 GeV Higgs-like boson~\cite{FGianotti,JIncandela,plb:2012gu,plb:2012gk}. However, there are strong hints suggesting possible existence of new physics at or beyond TeV scale, arising from, e.g., the compelling astrophysical evidences on dark matter, and the large hierarchy between electroweak and Planck scale. Thus searching for new physics beyond the SM remains both a theoretical and experimental pursuit.

One possible way to explore new phenomena in particle physics is to investigate bosonic anomalous couplings. Under the framework of SM, $SU(2)_L \times U(1)_Y$ gauge symmetry completely determines gauge boson interactions, while presence of any anomalous coupling vertex may generate observable deviation from SM prediction. Study on these vector-boson interactions, therefore, can either confirm the SM and the spontaneously symmetry breaking mechanism, or give hint on the form of new physics.
 
So far, explorations of anomalous trilinear gauge boson couplings (aTGCs) have already been carried out extensively at the LEP~\cite{Achard:2002vd,Abreu:2001rpa}, Tevatron~\cite{Gounder:1999wq,Abbott:1999aj}, and later at the LHC~\cite{Aad:2011xj,Chatrchyan:2012bd}
via vector boson pair production, while less effort has been made on probing anomalous quartic gauge boson couplings (aQGCs). It is important to note that aQGCs, although involving more complicated event topology and may be less sensitive at high energy colliders, are not mere substitution of aTGCs but should be regarded as an independent way of uncovering new physics, since, for example, the exchange of heavy bosons can generate tree-level contribution to quartic coupling while its effect on trilinear vertex appears only at one-loop and is consequently suppressed~\cite{Belanger:1992qh,Eboli:0310141}. Historically, T. Han {\it et al.} in 1989 calculated the scattering cross sections of various triple gauge boson productions at $e^+e^-$ and later $p\bar p$ colliders~\cite{Han:1989,Han:1995}. Monte Carlo (MC) studies were also performed later by \'{E}boli {\it et al.} at $e\gamma$ colliders and $\gamma\gamma$ colliders through the processes $e\gamma \rightarrow VV'F$ ($V,V' = W,Z,\gamma$ and $F = e,\nu$) and $\gamma\gamma \rightarrow W^{+}W^{-}V$ ($V = Z,\gamma$), giving constraints on relevant aQGCs~\cite{Eboli:9306306,Eboli:9503432}. Further MC work on aQGCs during that period were performed at $e^+e^-$ collider, and can be found in $e.g.,$~\cite{Belanger:2000,Stirling:2000, Dawson:1996aw,Denner:2001vr}. Direct constraints from experiments came mainly from the LEP at CERN, through $W^+W^-\gamma$~\cite{Abbiendi:1999aa,Abdallah:2003xn,Acciarri:2000en}, $Z\gamma\gamma$~\cite{Acciarri:2000rz} and $\gamma\gamma\nu\bar{\nu}(q\bar{q})$~\cite{Abbiendi:2004bf} channels. Due to limitations of center of mass energy, LEP's constraints at 95$\%$ confidence level on anomalous coupling constant are approximately at $10^{-2}$ GeV$^{-2}$ level, still two orders of magnitude larger than those from the oblique parameters $S$ and $U$ as argued in Ref.~\cite{Eboli:0009262} ($1 \times 10^{-4}\  \text{GeV}^{\text{-2}}$).

It is expected, comparatively, that the operation and its proposed upgrade within next few years of the Large Hadron Collider (LHC) at CERN will set more strict constraints on aQGCs. As shown in e.g., ~\cite{Eboli:0310141,Eboli:0009262,Royon:2010tw}, LHC can reach limits at about $10^{-5} - 10^{-6}$ $\text{GeV}^{\text{-2}}$ on the aQGCs, via the channel $W\gamma\gamma$, vector boson fusion (VBF) production of $\gamma\gamma$, $Z\gamma$ and $WW$. A more elaborated research by D. Yang {\it et al.} on \wwa production with full leptonic decay also confirmed the potential of LHC on probing \wwaa aQGCs~\cite{Yang:2012vv}.

In the paper, we are interested in measuring $W^{\pm}Z\gamma$ final states with full leptonic decay at the $\sqrt{s}=14$ TeV LHC and probing \wwza anomalous coupling. This work extends our previous study on \wwa and \wwaa aQGCs measurement~\cite{Yang:2012vv} as a further independent examination on tripe gauge boson physics at the LHC. Moreover, we believe \wza process has additional advantages as following: (1) \wza process suffers from less background due to the requirement of a leptonically decayed $Z$-boson reconstructed; (2) Being sensitive to \wwza vertex exclusively, \wza serves as a direct examination on \wwza aQGC.

Our paper is organized as follows. In Sec.~\ref{effwwza} we describe the photonic aQGCs effective Lagrangian and a novel parametrization to genuine \wwza aQGC. This is then followed by Sec.~\ref{eventsim}, showing our MC simulation framework and event selection details. Subsequently, Sec.~\ref{numresults} features numerical results, including the LHC sensitivities on \wza production with pure leptonic decays and the \wwza aQGC. Finally, we conclude in Sec.~\ref{discuss}.

\section{Effective Lagrangian for Photonic aQGCs}
\label{effwwza}
 
The quartic interaction can be constructed in a model-independent way with respect to the chiral Lagrangian approach~\cite{Eboli:0310141,Belanger:2000}. Assuming that new physics beyond the SM keeps $SU(2)_{L} \otimes U(1)_{Y}$ gauge invariance and $SU(2)_{c}$ custodial symmetry, we may write down the lowest order genuine aQGC \wwza operators in the form of independent Lorentz stuctures~\cite{Eboli:0310141,Belanger:2000,Yang:2012vv}. We list below two previously commonly used expressions of \wwza effective Lagrangian, both of which will be studied in this paper:
\begin{itemize}{
\item (\lowercase \expandafter {\romannumeral 1}) \emph{$\cal{C}\cal{P}$-violating Lagrangian} 
}\end{itemize}
\begin{equation}
{\cal L}_{n} = i\frac{\pi \alpha}{4 \Lambda^{2}} a_{n} \epsilon_{ijk} W^{(i)}_{\mu \alpha} W^{(j)}_{\nu} W^{(k)\alpha} F^{\mu \nu},
\label{anL}
\end{equation}
where $\alpha$ is the electroweak coupling constant, $a_{n}$ characterizes the strength of anomalous coupling, $\Lambda$ stands for new physics scale, $V_{\mu\nu}$ represents the field strength tensor given by $\partial_{\mu}V_{\nu}-\partial_{\nu}V_{\mu}$, and $W^{(i)}_{\mu}$ is the $SU(2)$ weak isospin triplet:
\begin{equation}
\overrightarrow{W_{\mu}} = \left( \begin{array}{ccc}
\frac{1}{\sqrt{2}}(W^+ + W^-)_{\mu} \\
\frac{i}{\sqrt{2}}(W^+ - W^-)_{\mu} \\
W^3_{\mu}-\frac{g'}{g}B_{\mu} \end{array} \right),
\end{equation}
with $\theta_{W}$ symbolizing the Weinberg mixing angle. The expression was widely used in references, but B\'elanger {\it et al.}~\cite{Belanger:2000} argued that it essentially violates $\cal{C}\cal{P}$ symmetry. However we emphasize that $\cal{C}\cal{P}$ invariance is not in general required by the first principle, thus we keep this form of effective Lagrangian for the purpose of comparison with previous experimental and MC outcomes~\cite{Eboli:9306306,Stirling:2000,Denner:2001vr,Abbiendi:1999aa,Abdallah:2003xn,Acciarri:2000en}. 
\begin{itemize}{
\item (\lowercase \expandafter {\romannumeral 2}) \emph{$\cal{C}\cal{P}$-conserving Lagrangian} 
}\end{itemize}
Following the notation in Ref.~\cite{Eboli:0310141} (see Eq.~(5) therein), there are 14 effective photonic operators relevant to aQGCs, specified by 14 independent coupling parameters $k^{w,b,m}_{0,c},\ k^{w,m}_{1,2,3},\ k^b_{1,2}$. After recombining and sorting various terms into similar Lorentz structures, one can see that among them five are related to anomalous \wwza vertex:
\begin{equation}
{\cal W}^Z_{0}=-\frac{e^2 g^2}{\Lambda^2}F_{\mu\nu}Z^{\mu\nu}W^{+\alpha}W^-_{\alpha},
\end{equation}
\begin{equation}
{\cal W}^Z_{c}=-\frac{e^2 g^2}{2\Lambda^2}F_{\mu\nu}Z^{\mu\alpha}(W^{+\nu}W^-_{\alpha} + W^{-\nu}W^+_{\alpha}),
\end{equation}
\begin{equation}
{\cal W}^Z_{1}=-\frac{e g_Z g^2}{2\Lambda^2}F^{\mu\nu}(W^+_{\mu\nu}W^-_{\alpha}Z^{\alpha} + W^-_{\mu\nu}W^+_{\alpha}Z^{\alpha}),
\end{equation}
\begin{equation}
\label{WZ2}
{\cal W}^Z_{2}=-\frac{e g_Z g^2}{2\Lambda^2}F^{\mu\nu}(W^+_{\mu\alpha}W^{-\alpha}Z_{\nu} + W^-_{\mu\alpha}W^{+\alpha}Z_{\nu}),
\end{equation}
\begin{equation}
\label{WZ3}
{\cal W}^Z_{3}=-\frac{e g_Z g^2}{2\Lambda^2}F^{\mu\nu}(W^+_{\mu\alpha}W^-_{\nu}Z^{\alpha} + W^-_{\mu\alpha}W^+_{\nu}Z^{\alpha}),
\end{equation}
where $g_Z = e/s_wc_w$, $g = e/s_w$, and we adopt the abbreviated symbol $c_w=\cos\theta_W$, $s_w=\sin\theta_W$.

Accordingly, the effective interactions can be expressed by the above operators as
\begin{equation}
\label{aqgceff}
{\cal L}_{eff} = \sum_{i} k^W_i {\cal W}^Z_{i},
\end{equation}
where the coefficient parameters $k^W_i(i=0,c,1,2,3)$ can be written as
\begin{equation}
k^W_0 = \frac{c_w}{s_w}k^w_0 - \frac{s_w}{c_w}k^b_0 + c_{zw}k^m_0,
\end{equation}
\begin{equation}
k^W_c = \frac{c_w}{s_w}k^w_c - \frac{s_w}{c_w}k^b_c + c_{zw}k^m_c,
\end{equation}
\begin{equation}
k^W_j = k^w_j + \frac{1}{2}k^m_j, \ \ (j=1,2,3),
\end{equation}
as shown in~\cite{Eboli:0310141}. Here $c_{zw}=(c^2_w-s^2_w)/(2c_ws_w)$. 

One would expect these $k^W_i$s are correlated with those coupling constants that characterize \wwaa, $ZZ\gamma\gamma$ and $ZZZ\gamma$ interactions~\cite{Eboli:0310141}. A practicable way for decorrelation is to seek for proper subspace of these 14 parameters ($k^{w,b,m}_{0,c},\ k^{w,m}_{1,2,3},\ k^b_{1,2}$), namely, to impose extra restrictions on $k^i_j$, leaving only \wwza vertex non-vanishing. A simple parametrization is proposed in Ref.~\cite{Belanger:2000}, requiring $k^m_2=-k^m_3$ and others vanished, and then one has:
\begin{equation}
{\cal L}_{eff} = \frac{1}{2}k^m_2({\cal W}^Z_{2} - {\cal W}^Z_{3}).
\label{k2mL}
\end{equation}
One alternative solution involving four independent parameters $k^w_0,k^m_0,k^w_2$, and $k^m_2$ is given in Appendix.~\ref{appendixA}, which can be expressed as:
\begin{equation}
k^W_0=\frac{1}{c_ws_w}(k^w_0+\frac{1}{2}k^m_0),
\label{newpar1}
\end{equation}
\begin{equation}
k^W_c=\frac{1}{c_ws_w}(k^w_0+\frac{1}{2}k^m_0),
\end{equation}
\begin{equation}
k^W_1=-k^w_0-\frac{1}{2}k^m_0,
\end{equation}
\begin{equation}
k^W_2=k^w_2+\frac{1}{2}k^m_2,
\end{equation}
\begin{equation}
k^W_3=-(k^w_0+\frac{1}{2}k^m_0)-(k^w_2+\frac{1}{2}k^m_2).
\label{newpar5}
\end{equation}
The 4-dimensional solution automatically includes Eq.~(\ref{k2mL}), if we set other three parameters than $k^m_2$ to zero. In the following, we stick to this 4-dimensional parametrization, and in our analysis we first vary each parameter separately while setting others equal to zero, and then we also investigate the correlation of two individual parameters and draw contours at 95\% confidence level. 
 
Finally, we want to mention that, for sufficiently high energy collision, the effective Lagrangian leads to tree-level unitarity violation and is usually regulated by introducing appropriate form factor (ff) as following~\cite{Eboli:0310141}:
\begin{equation}
k^m_2 \rightarrow \frac{k^m_2}{(1+\hat{s}/\Lambda^2_u)^n}
\end{equation}
where $\hat{s}$ is the the partonic center-of-mass energy and $\Lambda_{u}$ represents the new physics scale. In Ref.~\cite{Yang:2012vv} we see that a reasonable choice of form factor can indeed satisfy the unitarity requirement. In this paper we choose $n=5$, $\Lambda_{u}=2.5$ or $\infty$ TeV. The latter choice of $\Lambda_u$ equals to no form factor at all.

\section{Event Simulation and Selection}
\label{eventsim}
We carry out our MC simulations within \mgme\ v5~\cite{Alwall:2007st,MadGraph:2012,Maltoni:2002qb}. The effective Lagrangian of \wwza aQGCs is implemented into \madgraph\ based on the FeynRules~\cite{Christensen:2008py}-UFO~\cite{Degrande:2011ua}-ALOHA~\cite{deAquino:2011ub} framework. The signal and background concerned are initially generated at parton level by \madgraph\ and \madevent, and are then passed through the interface to \pythia\ 6 for parton showering and hardronization~\cite{Sjostrand:2003wg}. The detector simulations are then done using \delphes\ 2.0~\cite{Ovyn:2009tx} package, where we focus on the CMS detector at the LHC. Finally, all events are delivered to {\sc ExRootAnalysis}~\cite{ExRootAnalysis} and analyzed with ROOT~\cite{ROOT}. The work flow has also been used in our previous studies~\cite{Yang:2012vv, Liu:2012rv}.

\begin{figure}[ht]{
\centering
\subfigure[TGC: type 1]{
    \label{diagram:a}
    \includegraphics[width=0.35\textwidth]{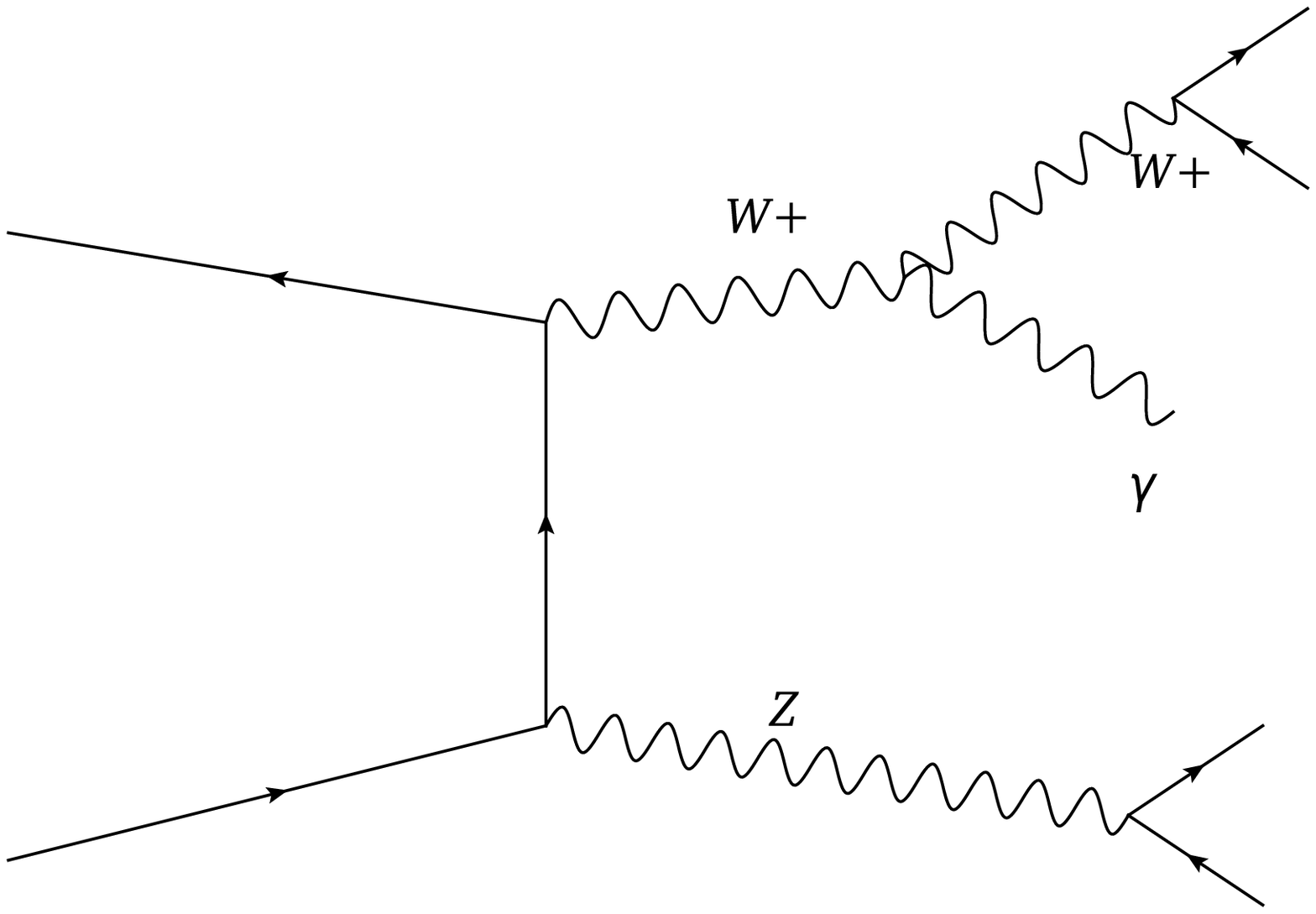}
}
\subfigure[TGC: type 2]{
    \label{diagram:b}
    \includegraphics[width=0.35\textwidth]{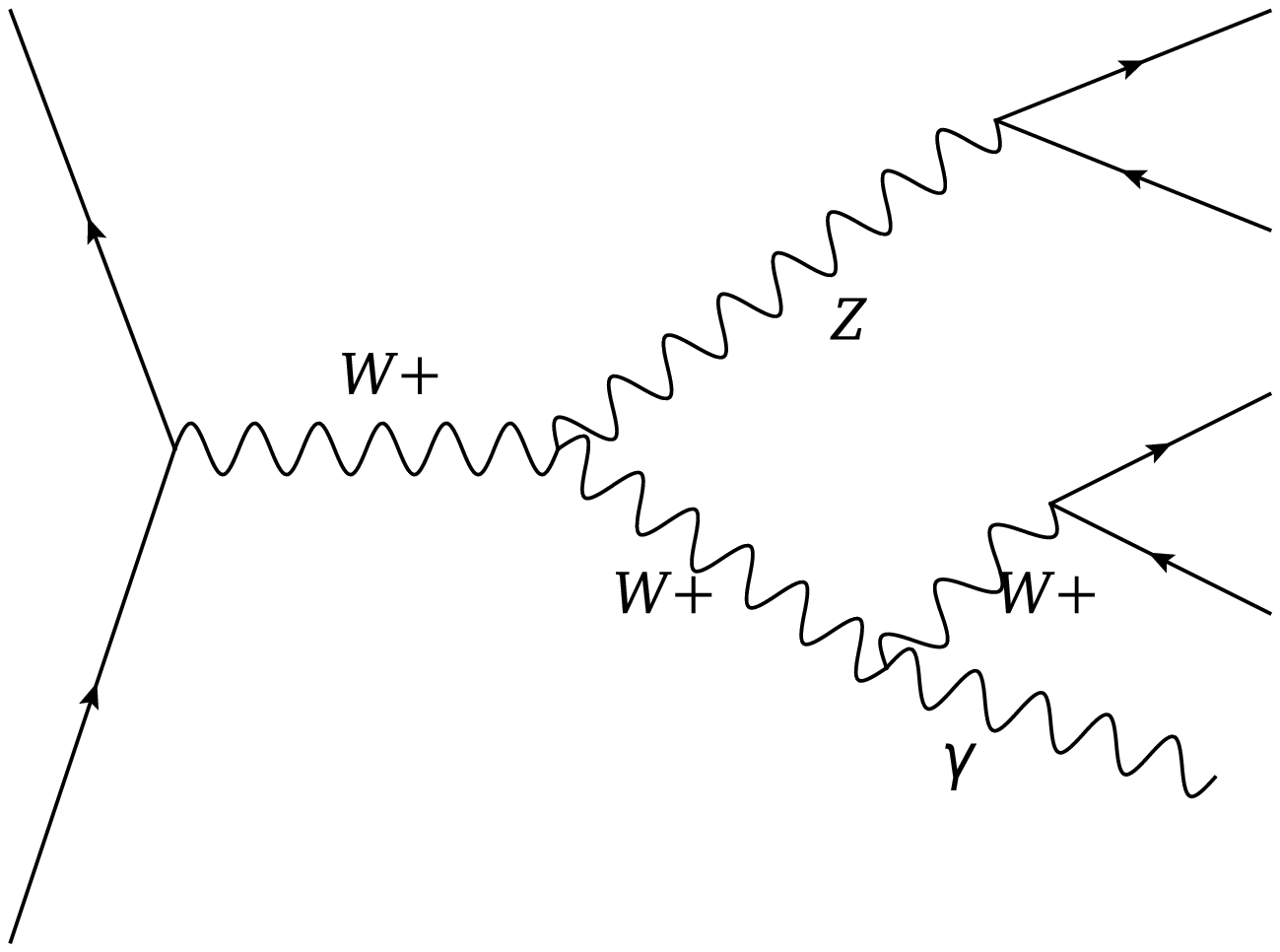}
}
\subfigure[(Anomalous) QGC]{ 
    \label{diagram:c}
    \includegraphics[width=0.35\textwidth]{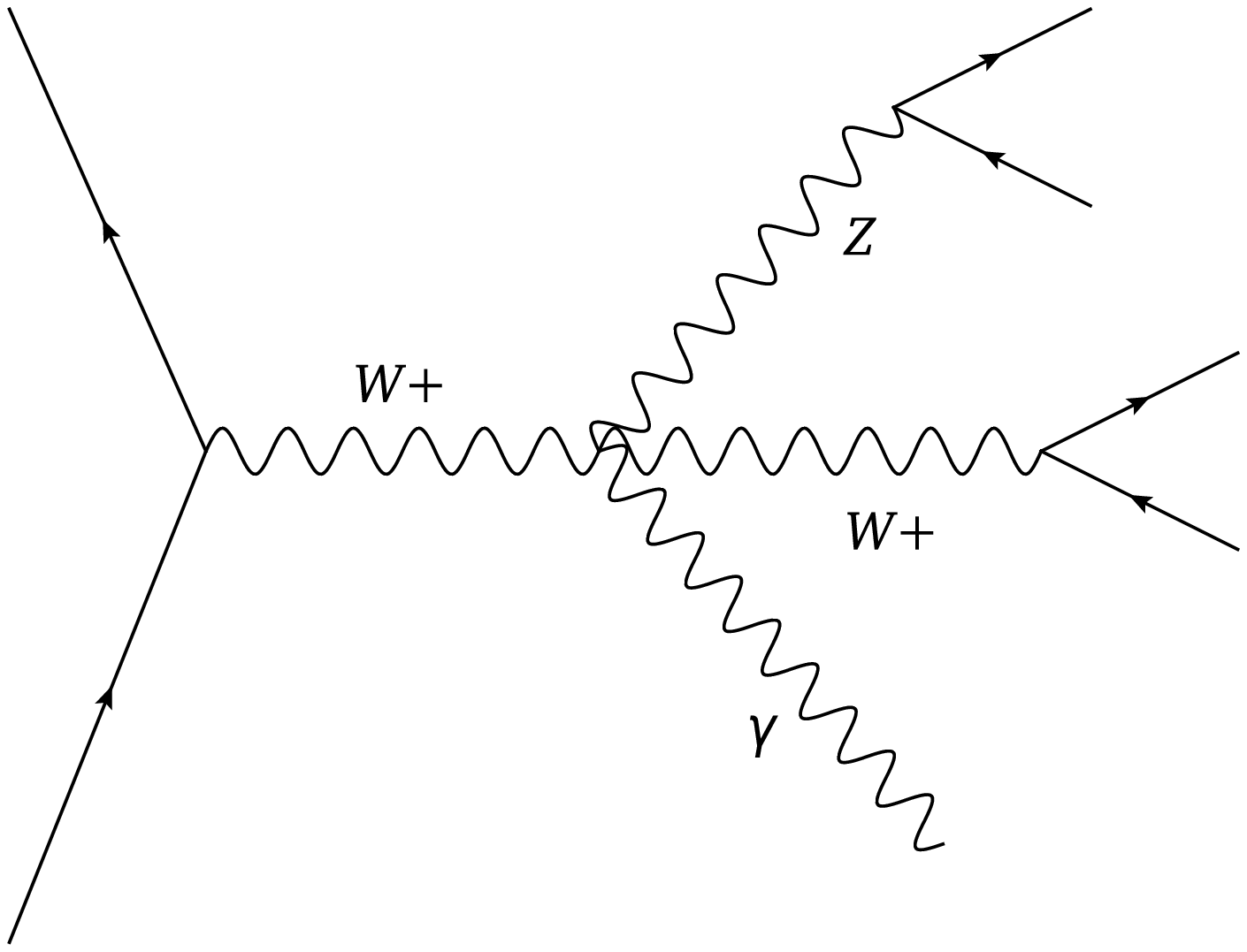}
}
\subfigure[QED Radiations from $WZ$]{ 
    \label{diagram:d}
    \includegraphics[width=0.35\textwidth]{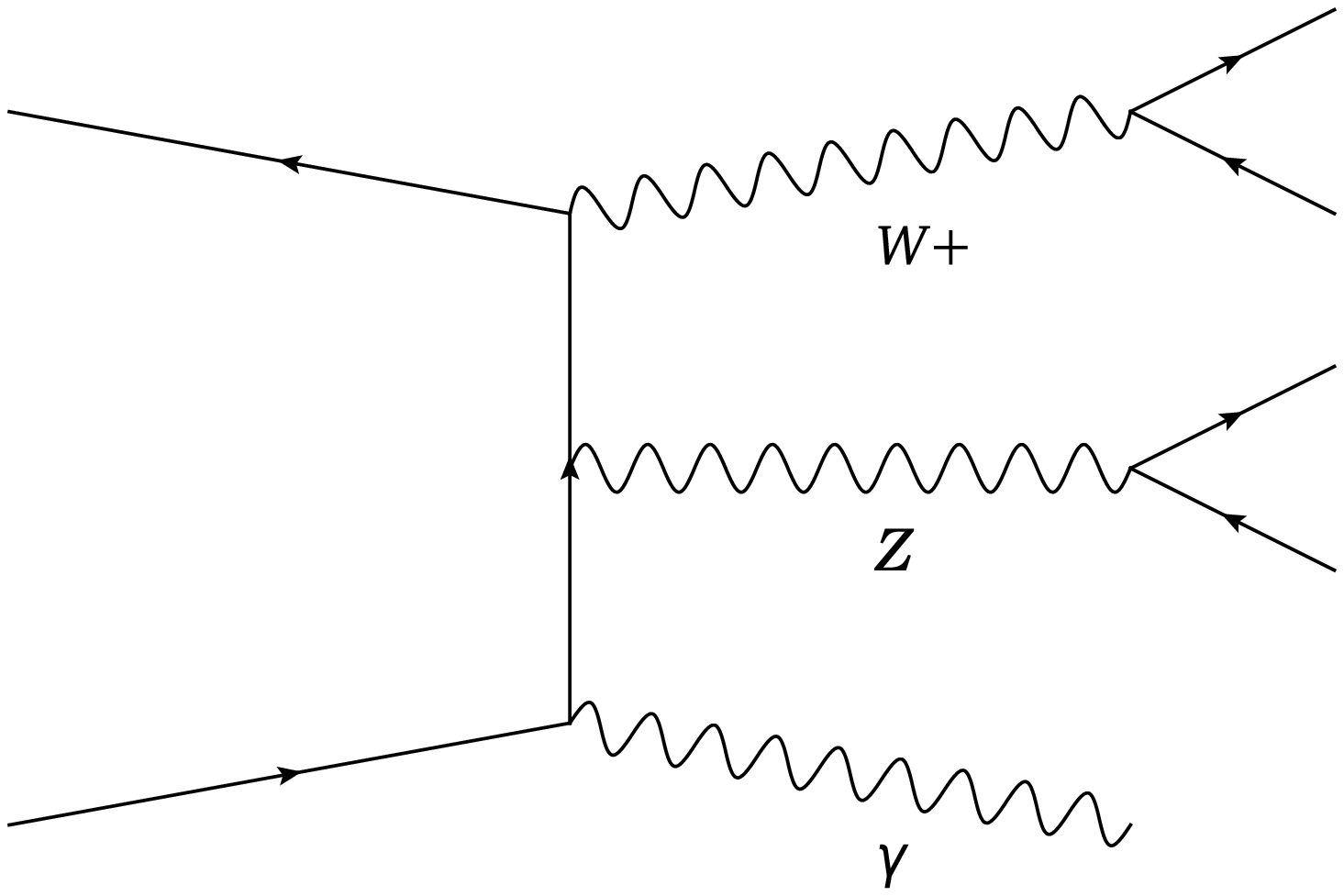}
}
\caption{\label{FeynDiagram} Sample Feynman diagrams that contribute to \wza productions in $p\ p$ collision.}
}
\end{figure}

The characteristic signal we are interested in contains three well-defined leptons with total electric charge $\pm1$, in association with large missing transverse energy \met. Besides, there is one and only one pair of oppositely charged lepton with same flavor originated from $Z$ boson decay. Some example Feynman diagrams are plotted in Fig.~\ref{FeynDiagram}, for \wza production at the LHC,
in the di-leptontic final state $l\nu L\bar{L} \gamma$, with $l,L=e,\mu$ and $\tau$. Note $\tau$ decays into $e, \mu$ at the ratio of about 35\% and is handled with \tauola~\cite{tauola}.

In Fig.~\ref{FeynDiagram}, two types of diagrams, Figs.~\ref{diagram:a} and \ref{diagram:b}, involve TGCs and are not sensitive to aQGCs, while Fig.~\ref{diagram:d} can also be seen as the initial and final state radiations (ISR and FSR) from the $WZ$ production process, generated by \pythia. However, the ISR and FSR approximations in \pythia\, should break down for hard or wide scattering photon, e.g., when the transverse momentum of $\gamma$, $P_{T\,\gamma}$ is large. Note also this subset of contributions to \wza is not related to QGCs, thus it would be interesting and important to show the overall \wza results subtracting the contributions of the ISR/FSR approximations of Fig.~\ref{diagram:c}, which we denote as $\rm{pure\_Vs}$:
\begin{equation}
\label{pureVs}
{\rm pure\_Vs} \equiv W^{\pm}Z\gamma - W^{\pm}Z\ {\rm ISR/FSR}.
\end{equation}

Aside from $WZ$ ISR/FSR, another six backgrounds are taken into account: $ZZ$, $ZZ\gamma$, $ZZZ$, $WWW$, $WWZ$ and $t\bar{t}Z$. Note that multi-lepton ($n>3$) final state can be possible backgrounds with additional leptons misidentified. Here we do not consider backgrounds with photons from jet fragmentation, in which the photons tend to be close to jets and the contributions can be suppressed efficiently via photon isolation cuts (see e.g. Ref.~\cite{Frixione:1998}).

We choose the following pre-selection cuts to generate unweighted events at parton level
with \mgme\, to interface later with \pythia\, and \delphes\,, 
\begin{itemize}{
\item  (1) $P_{T\,\gamma, l} \geq 15 $ GeV,
\item  (2) $|\eta_\gamma| < 2.5, |\eta_l| < 2.5$,
\item  (3) $R_{ll} > 0.4, R_{l\gamma} > 0.4$,
}\end{itemize}
where $R$ denotes the separation $\sqrt{\Delta\phi^2 + \Delta\eta^2}$ in which $\phi$ being the azimuthal angle and $\eta$ the pseudo-rapidity of a particle. Note, however, for the backgrounds involving misidentified leptons, we do not require any of the above cuts on leptons in order not to make bias.

Moreover, in the hard process generation with \mgme\, we adopt the CTEQ6L1 parton distribution functions
(PDFs)~\cite{Pumplin:2002vw} and set the renormalization and factorization scales as the transverse mass of the core process.

Further reconstruction cuts are then imposed on the reconstructed objects in the \delphes\, settings cards,
\begin{itemize}{
\item $P_{T\,e,\mu,\gamma} \geq 15\,$GeV, and $|\eta_{e,\mu,\gamma}|<2.4$.  
\item Jets are clustered according to the anti$-k_t$ algorithm with a cone radius $\Delta R = 0.5$. 
Moreover, $P_{T,j}>P^{cut}_{T,j}$ (25\,GeV by default) and $|\eta_{j}|<5$ are required.
}\end{itemize}

Tighter cuts are set in the analysis steps posterior to detector simulation, 
\begin{itemize}{
\item  (1) The leading photon $P_{T,\gamma}$ should exceed the threshold, $P_{T,\gamma}^{cut}$, the value of which will be optimized,
\item  (2) $\met > 40$ GeV,
\item  (3) To distinguish from background processes (e.g., $t\bar{t}Z$) with more hard jets, require $P_{T,j} < P_{T,j}^{veto}$ and at most 1 jet is allowed,
\item  (4) Three and only three leptons with $\pm1$ total charge,
\item  (5) Exclude $b$-tagged jets to suppress top quark-related production,
\item  (6) $R_{j\gamma}, R_{l\gamma}, R_{ll} > 0.5$,
\item  (7) One and only one pair of oppositely charged lepton with same flavor comes from $Z$ boson decay, with $|m_{ll}-M_Z| < 10$ GeV.
}\end{itemize}

In \delphes, photons and charged leptons may overlap with the jet collections: \delphes\, first reconstructs photons and leptons based on MC information, and then jets which can be seeded from the already reconstructed photons or leptons. In our analysis, we clean the lepton collections from jets by requiring the \delphes's calculated ``EhadOverEem" (the energy deposition in the Hadron Calorimeter over the one in the Electromagnetic Calorimeter) smaller than 1. Moreover, we remove any jet with $R_{j \gamma}<0.001$ as it would be indeed most like a photon.

\section{Numerical Results}
\label{numresults}
\subsection{\wza Production}

As a first step, we are interested in estimating the feasibility of observing triple gauge boson \wza production at the LHC, before going into aQGCs. As mentioned before, we are also interested in comparing overall \wza results with the ones subtracting ISR/FSR contributions from $WZ$ processes (see Eq.~(\ref{pureVs})).

To optimize our results, we introduce further the following 3 requirements (similar as in our previous \wwa study~\cite{Yang:2012vv}), in addition to all the cuts as mentioned in Sec.~\ref{eventsim}:
(A) Maximize sensitivities by varying photon $P_T$ threshold cut $P_{T,\gamma}^{cut}$; (B) Keep the best $P_{T,\gamma}^{cut*}$, optimize over $P_{T,j}^{veto}$; (C) Keep the best $P_{T,\gamma}^{cut*}$ and $P_{T,j}^{veto*}$ values, vary $P_{T,j}$ threshold cut, and further require that within the interval $[P_{T,j}^{cut*}, P_{T,j}^{veto*}]$ at most 1 jet exists. The significance is defined by~\cite{stat:atlas}
\begin{equation}
\label{sig}
Q=(1+\frac{N_s}{N_b})^{N_{obs}}e^{-N_s},\ {\rm significance}=\sqrt{2{\rm ln}Q},
\end{equation}
where $N_s$, $N_b$ stand for number of signal and number of backgrounds, and $N_{obs}=N_s+N_b$.

\begin{figure}[ht]
\centering
\subfigure[]{
    \label{PTa_cut}
    \includegraphics[width=0.40\textwidth]{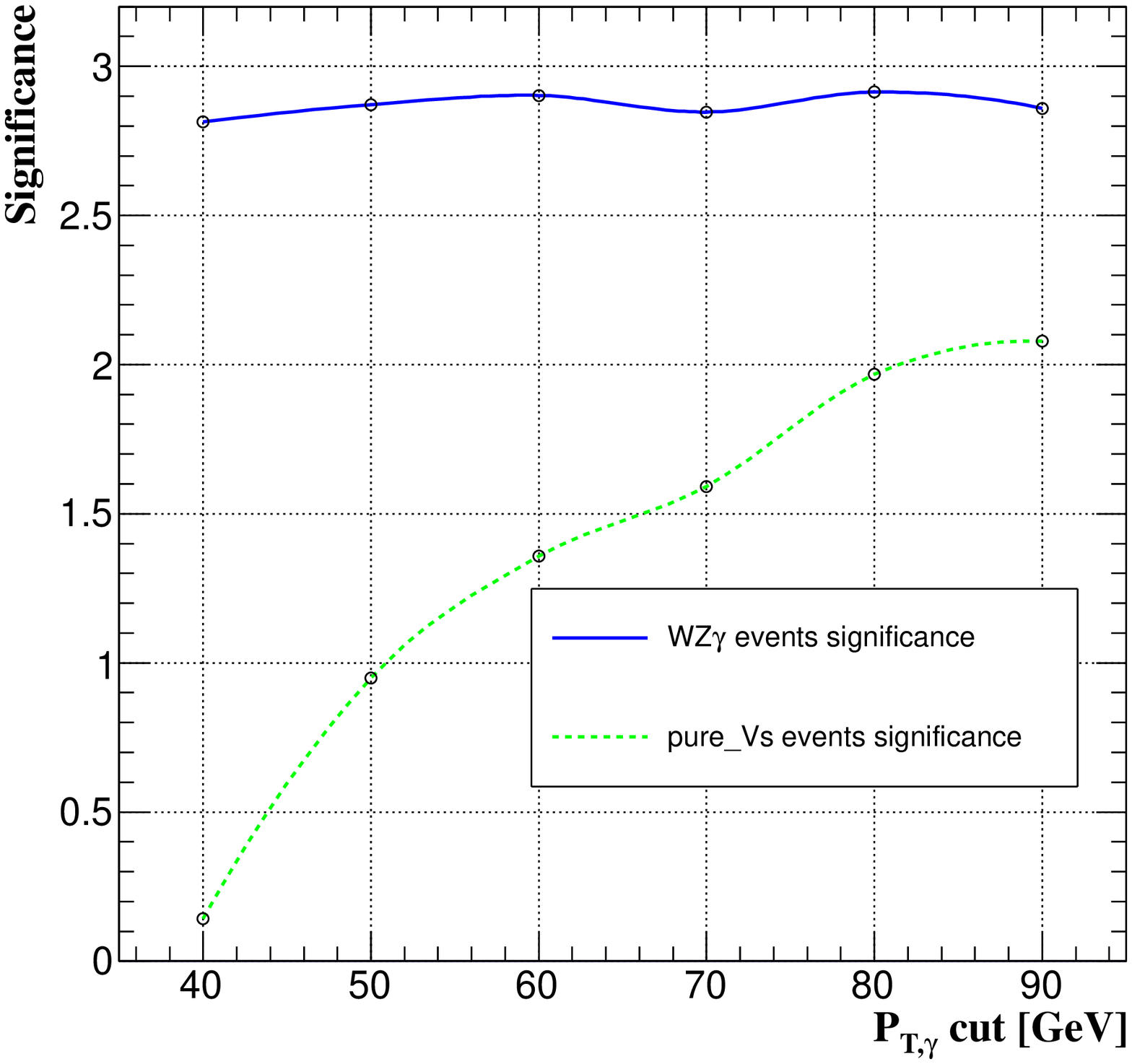}
}
\subfigure[]{
    \label{PTj_veto}
    \includegraphics[width=0.40\textwidth]{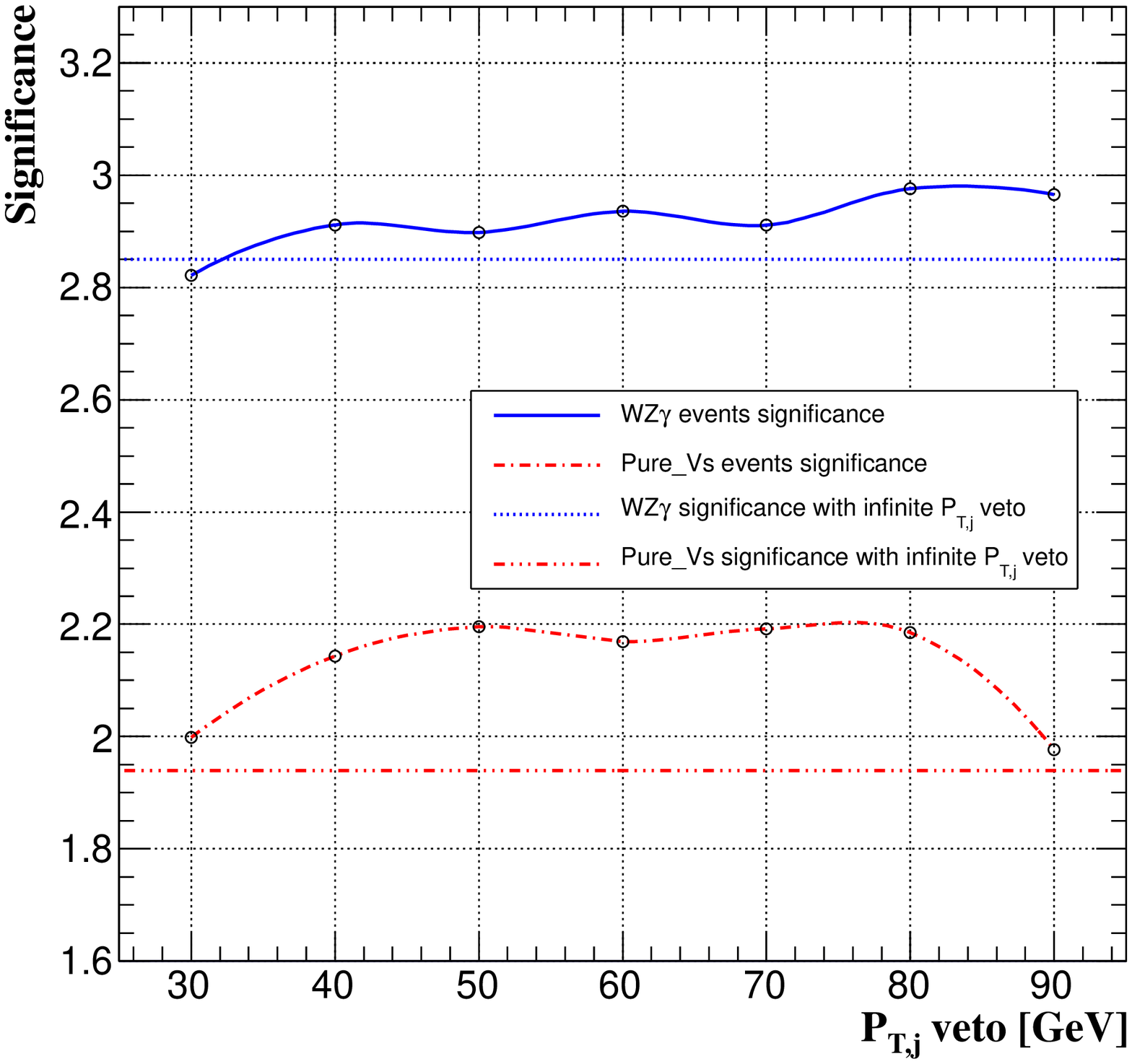}
}
\subfigure[]{ 
    \label{Numjet}
    \includegraphics[width=0.40\textwidth]{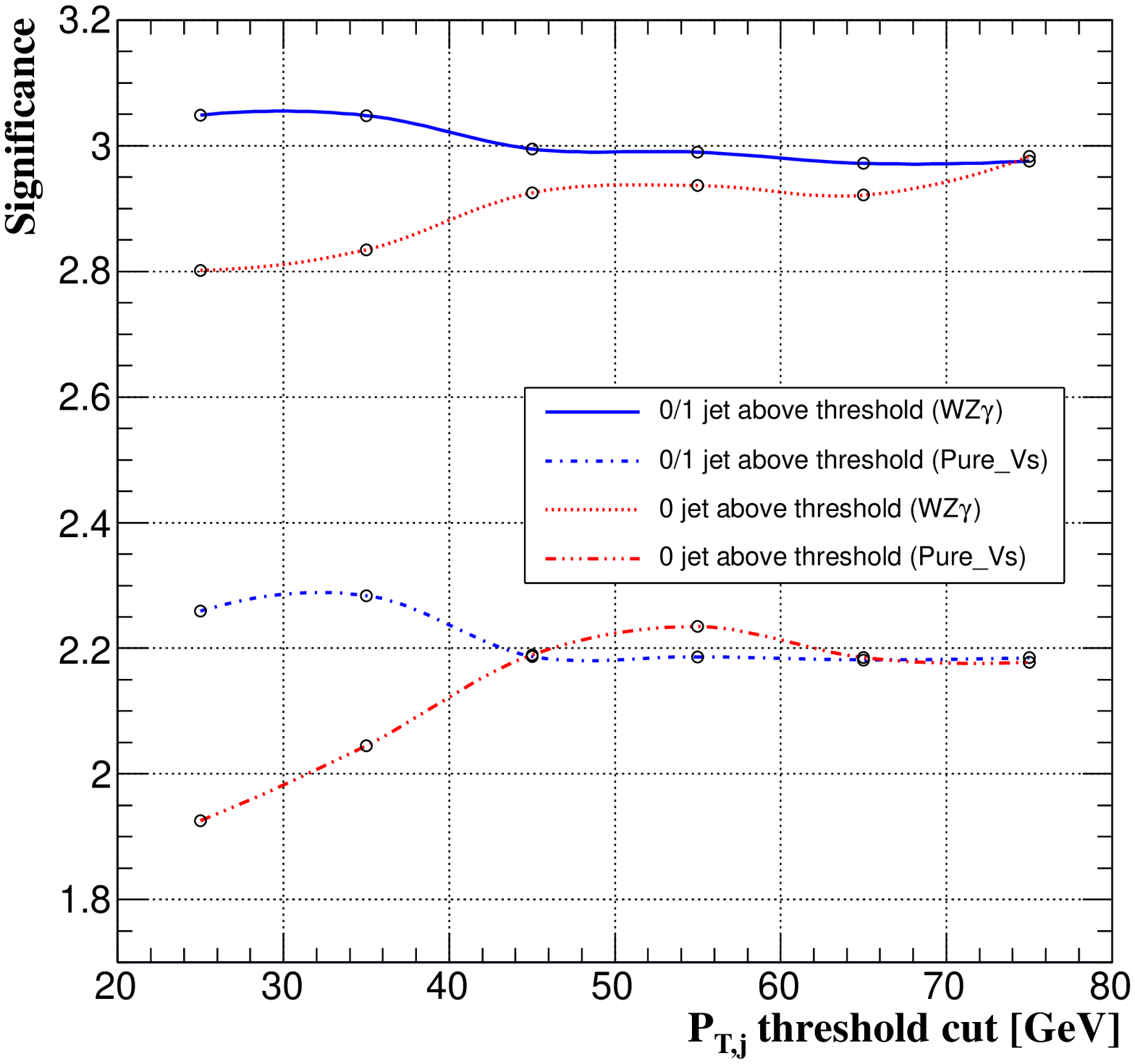}
}
\caption{\label{CutSelection} 
\wza and $\rm{pure\_Vs}$ significances, varying (a) $P_{T,\gamma}^{cut}$; (b) $P_{T,j}^{veto}$ in the presence of optimized $P_{T,\gamma}^{cut*}$; (c) $P_{T,j}^{cut}$ with optimized $P_{T,\gamma}^{cut*}$ and $P_{T,j}^{veto*}$.}
\end{figure}

We list the event numbers for the signal and backgrounds in Table~\ref{tab}, with the optimized parameters (optimized for $\rm{pure\_Vs}$ contributions) from the above 3 steps: ($A^{\ast}$) $P_{T,\gamma}^{cut*}=80\,$GeV, ($B^{\ast}$) $P_{T,j}^{veto*}=80\,$GeV, ($C^{\ast}$) $P_{T,j}^{cut*}=35\,$GeV.
Related K-factors for the signal and backgrounds are also listed with references in Table~\ref{tab}. Correspondingly, the significances are shown in Fig.~\ref{CutSelection}, calculated with Eq.~(\ref{sig}).  

\begin{center}
\begin{table*}[h!]
\begin{tabular}{|c||c|c||c|c|c|}
\hline
\multirow{2}{*}{Processes} & $\sigma$ (LO) & K-factor
& \multicolumn{3}{c}{ Events } \vline\\
\cline{4-6}
& [fb] & [Ref.] & {\tiny($A^{\ast}$) $P^{cut*}_{T,\gamma}=80$ GeV} & {\tiny($B^{\ast}$) $P^{veto*}_{T,j}=80$\,GeV} & {\tiny($C^{\ast}$) $n_j=0,1$, $P_{T,j}^{cut*}=35$ GeV}\\\hline 
\wza              & 0.89     &  2.0~\cite{K-wza}     &     3.78      &     3.48            &     3.41            \\\hline 
I(F)SR $WZ$          & 349.4     &  1.8~\cite{K-wz}       &    0.76      &     0.50           &     0.44            \\\hline 
$ZZ\gamma$         & 0.24     &  1.4~\cite{K-zza}      &     0.19      &     0.18           &     0.17            \\\hline 
$ZZ$        & 99.4     &  1.6~\cite{K-wz}     &     0.16      &     0.16          &     0.16            \\\hline 
$ZZZ$  & 0.059     &  1.5~\cite{K-zzz}     &     0.008      &     0.007           &     0.007            \\\hline 
$WWW$  & 1.72     &  1.8~\cite{K-zzz}  &     0  &   0   &   0 \\\hline 
$WWZ$  & 0.96     &  1.9~\cite{K-wwz}          &     0.085      &     0.079           &     0.073            \\\hline 
$t\bar{t}Z$  & 6.16     &  1.4~\cite{K-ttz}          &     0.35      &     0.16           &     0.086            \\\hline 
\end{tabular}
\caption{\label{tab} Cut flow at the LHC with $\sqrt{s}=14$ TeV and integrated luminosity of $100\fbinv$.}
\end{table*}
\end{center}

More details can also be checked in Fig.~\ref{CutSelection} for cuts optimization. Note we also give the $\rm{pure\_Vs}$-curves to show the results after subtracting ISR/FSR contributions from $WZ$ processes, as mentioned above. One can see that higher $P_{T,\gamma}^{cut}$ only slightly changes the $WZ\gamma$ significance as it removes both signal and backgrounds in a similar way, on the other hand, pure aQGCs  $\rm{pure\_Vs}$-significance is enhanced quickly as more ISR/FSR background is killed. $P_{T,j}^{veto}$ cuts a bit more top-related backgrounds but the overall effects on significance is small. Increasing jet reconstructing cut $P^{cut}_{T,j}$, increases the 0-jet contributions while decreases the 1-jet ones, as expected. The overall 0+1 jet significances shrink slightly, as signal events are also discarded.

Above all, a significance about 3$\sigma$ can be achieved to observe $WZ\gamma$ production at the 14 TeV LHC, and does not depend so much on the cuts as mentioned above. Note a large portion of $WZ\gamma$ events can come from the QED ISR/FSR $WZ$ which is not related to QGCs, as shown by the $\rm{pure\_Vs}$-curves in Fig.~\ref{CutSelection}, however, sticking to large $P_{T\,\gamma}$ lower cut ($\sim 80$\,GeV), one can still get a total significance about 2$\sigma$ from $\rm{pure\_Vs}$ contributions.

\subsection{Anomalous \wwza Couplings}
\label{wwzaaQGC}
\begin{figure}[ht]
\centering
\subfigure[]{
    \label{k2mPhoton}
    \includegraphics[width=0.40\textwidth]{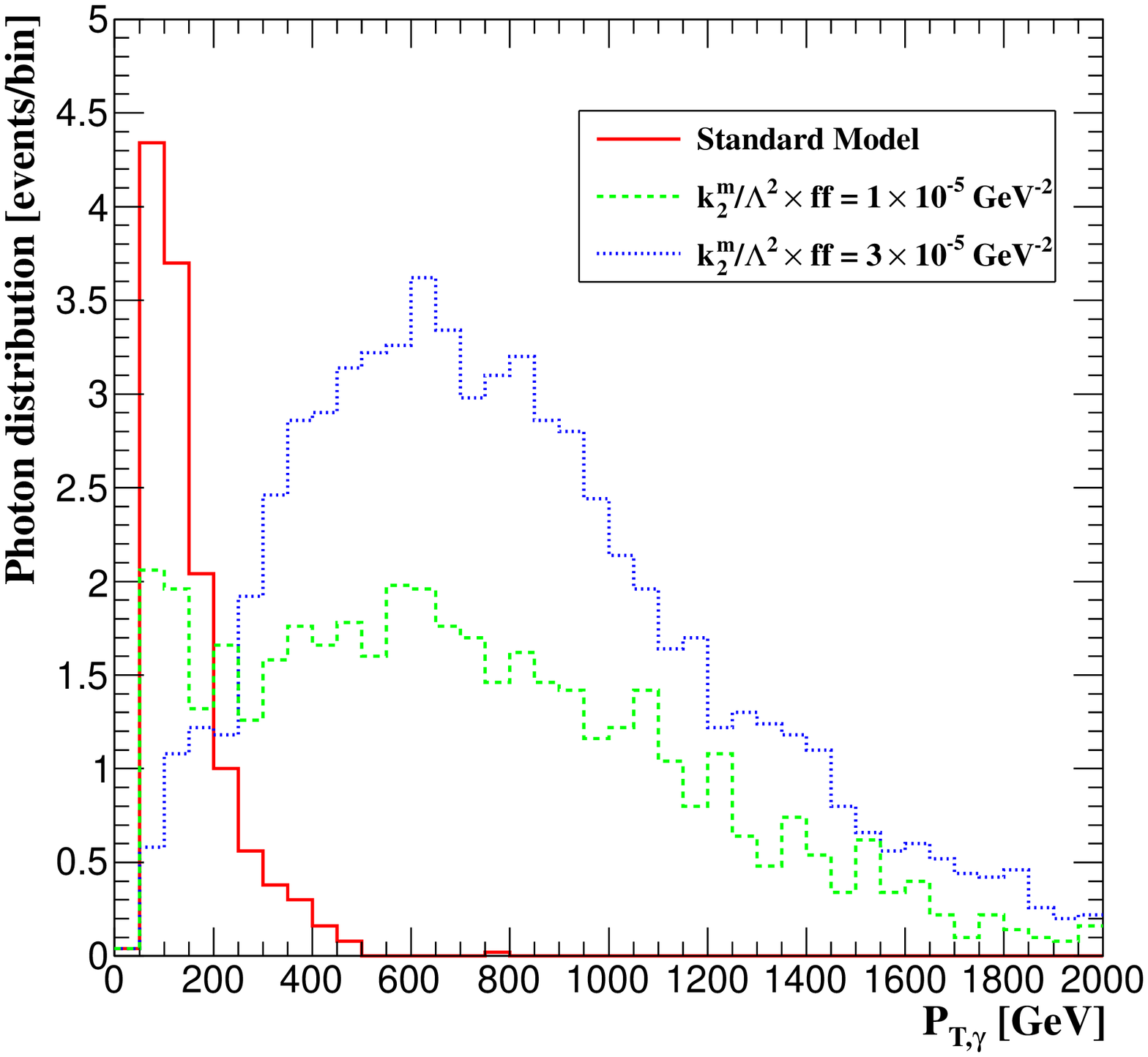}
}
\subfigure[]{
    \label{k2mLepton}
    \includegraphics[width=0.40\textwidth]{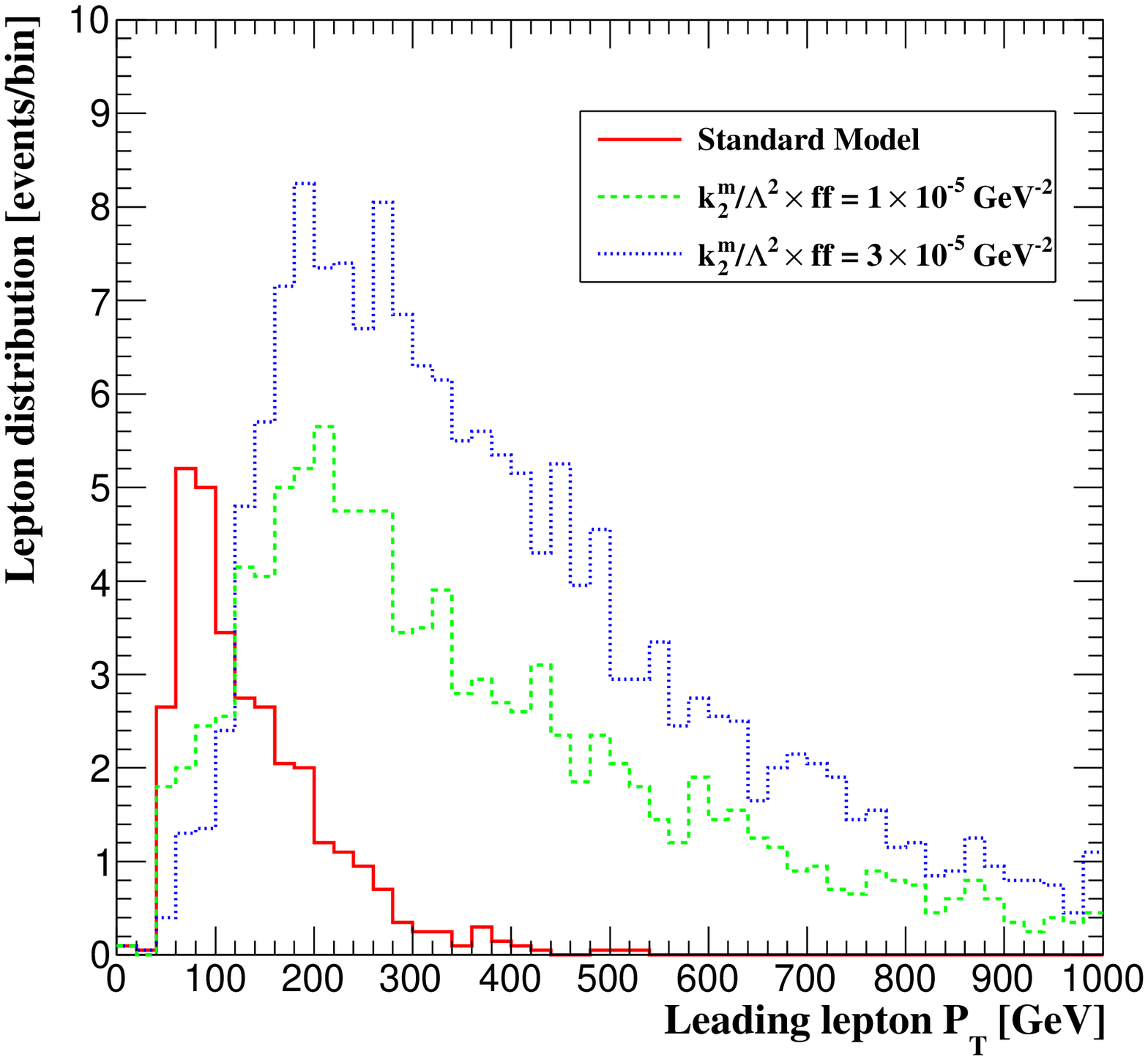}
}
\subfigure[]{ 
    \label{anPhoton}
    \includegraphics[width=0.40\textwidth]{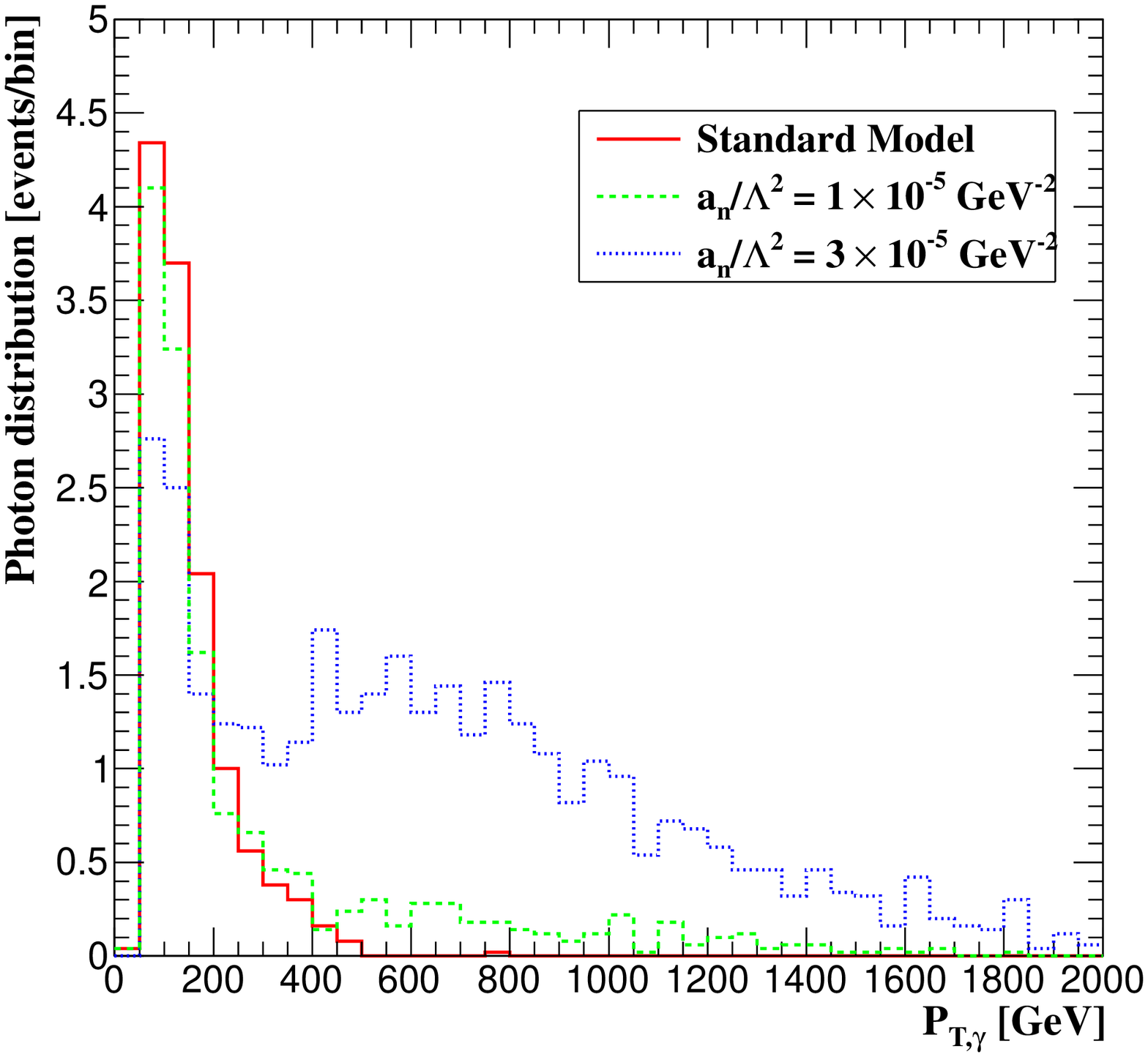}
}
\subfigure[]{ 
    \label{anLepton}
    \includegraphics[width=0.40\textwidth]{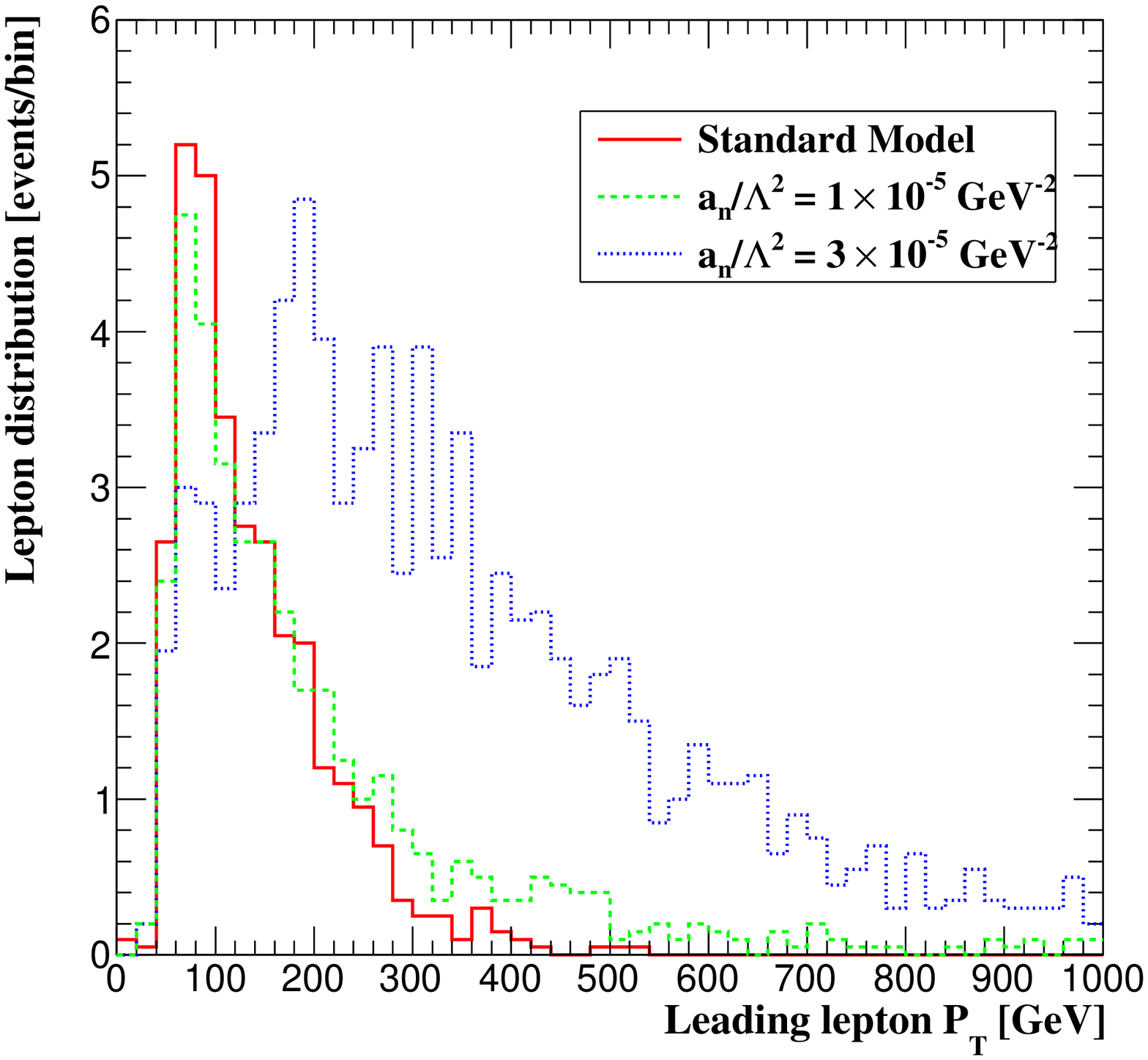}   
}
\caption{\label{CutRefine} 
Comparisons of the differential distributions for $WZ\gamma$ productions at the LHC in leading lepton $P_T$ and photon $P_T$, with \wwza aQGCs following
(a-b) ${\cal CP}$ conserving Lagrangian (Eq.~(\ref{k2mL})); (c-d) ${\cal CP}$ violating Lagrangian (Eq.~(\ref{anL})) without form factor.
}
\end{figure}

The \wza signal process can be sensitive to aQGCs \wwza. As shown in Fig.~\ref{CutRefine}, the aQGCs lead to excesses on the hard tails in various kinematic region. One thus can refine the cuts in Sec.~\ref{eventsim} to enhance the sensitivity to aQGCs as following, e.g. :

\begin{itemize}{
\item (1) $P_{T,\gamma} \geqslant 200$ GeV,
\item  (2) $P_{T,l} \geqslant 120$ GeV.
}\end{itemize}

\begin{figure}[ht]
\centering
\subfigure[]{
    \label{K2M}
    \includegraphics[width=0.40\textwidth]{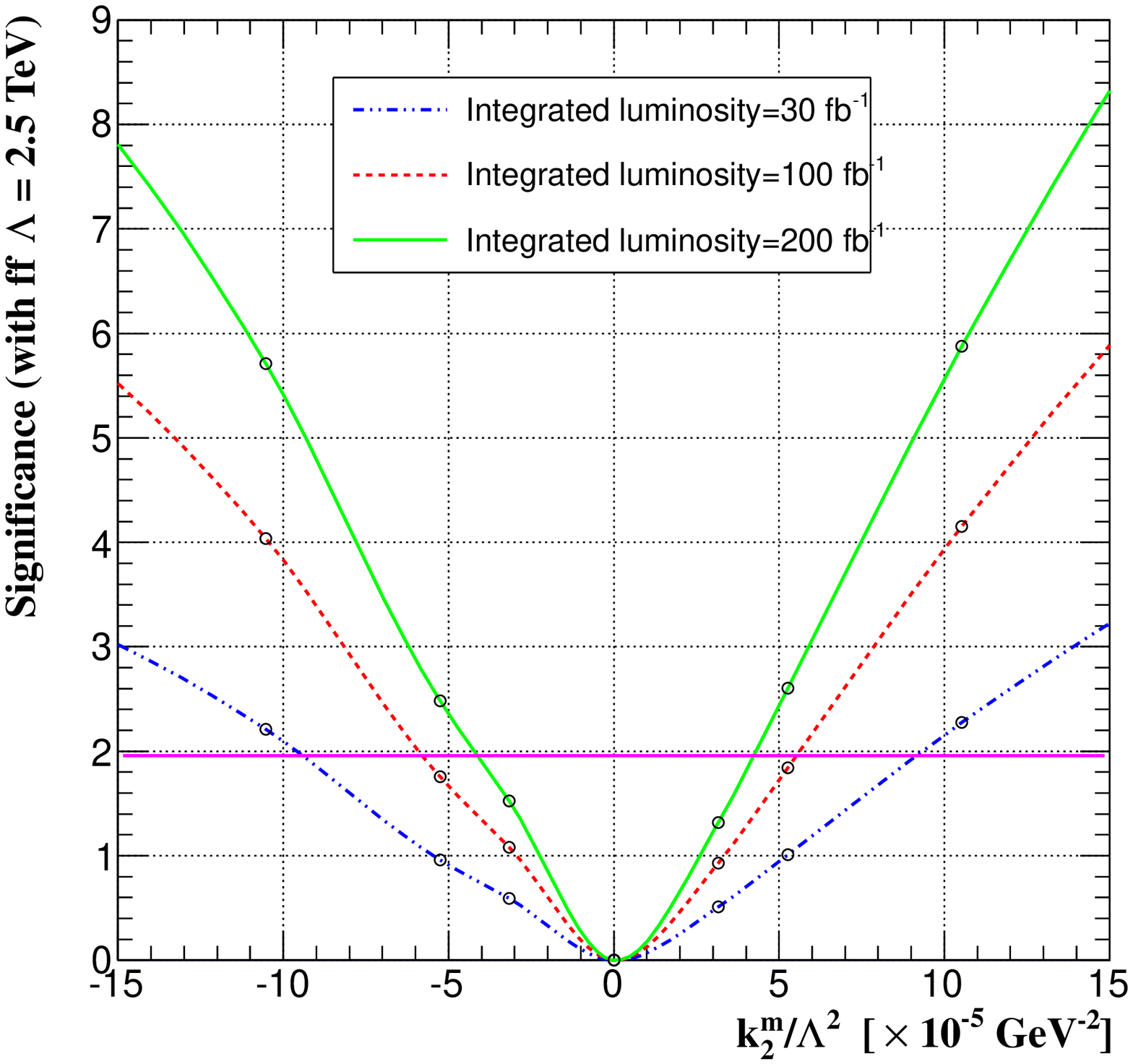}
}
\subfigure[]{
    \label{nffK2M}
    \includegraphics[width=0.40\textwidth]{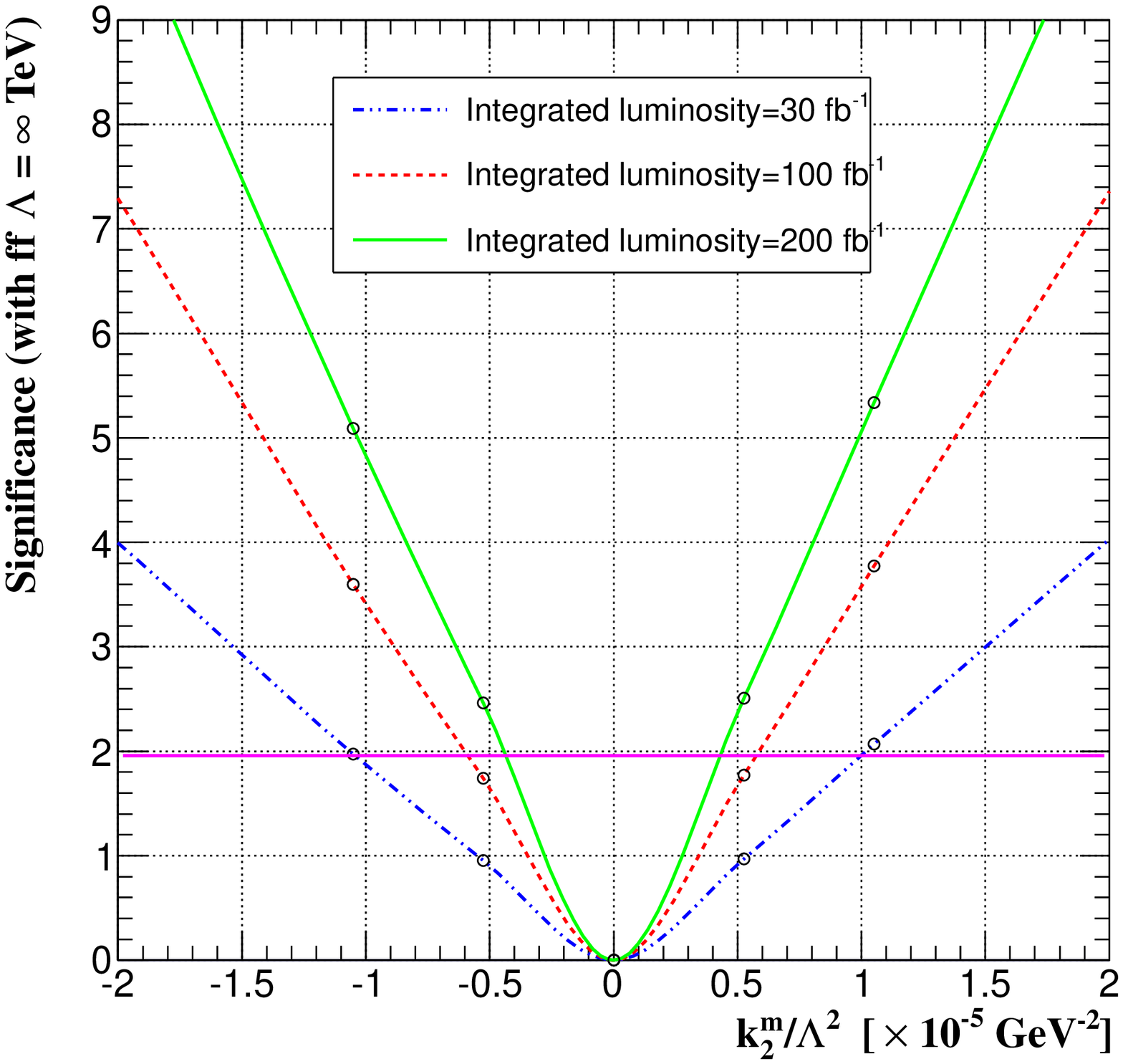}
}
\subfigure[]{
    \label{AN}
    \includegraphics[width=0.40\textwidth]{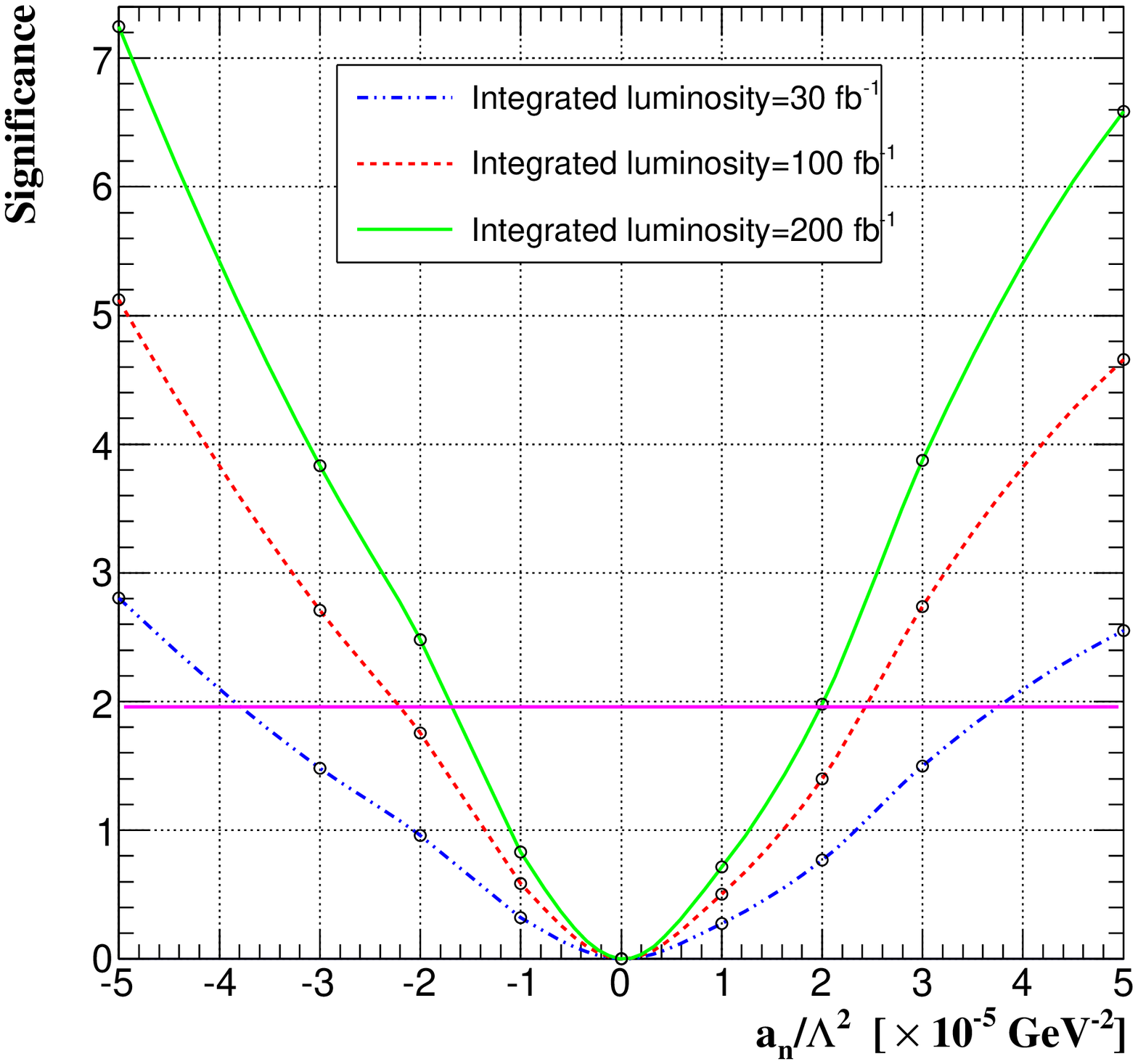}
}
\caption{\label{Limits} Anomalous coupling signal significance: (a) ${\cal CP}$ conserving Lagrangian (Eq.~(\ref{k2mL})) with form factor n=5 and $\Lambda_u=2.5$ TeV; (b) ${\cal CP}$ conserving Lagrangian (Eq.~(\ref{k2mL})) with form factor n=5 and $\Lambda_u=\infty$ TeV;(c) ${\cal CP}$ violating Lagrangian (Eq.~(\ref{anL})) without form factor.
}
\end{figure}

After all these selection cuts, the significances are calculated and displayed in Fig.~\ref{Limits} as functions of the \wwza aQGCs within ${\cal CP}$ conserving Lagrangian (Eq.~(\ref{k2mL})) or ${\cal CP}$ violating Lagrangian (Eq.~(\ref{anL})), at the 14 TeV LHC, with an integrated luminosity of 30, 100 and 200 \fbinv, respectively. The horizontal dash lines here correspond to the 95\% confidence level limit. Note here the signal is defined as $({\rm QGCs-SM}\,WZ\gamma)$.

Performing linear interpolation, 95\% C.L. limits on \wwza aQGCs are obtained (the constraints of $k_2^m, k^w_0, k^m_0$ and $k^w_2$ are derived assuming independence among them, $i.e.$, varying one parameter while setting others zero, with the parametrization given in Appendix.~\ref{appendixA}) at the 14 TeV LHC, with an integrated luminosity of 30(100)[200] \fbinv, respectively:

\begin{itemize}
\item{(1) $\Lambda_u=2.5$ TeV\\
\ $-9.5 (-5.7) [-4.1] \times 10^{-5}$ $< k_2^m/\Lambda^2 <$ $9.2 (5.5) [4.2] \times 10^{-5}$ GeV$^{-2}$,\\
\ $-2.6 (-1.5) [-1.0] \times 10^{-5}$ $< k_0^w/\Lambda^2 <$ $2.3 (1.4) [0.9] \times 10^{-5}$ GeV$^{-2}$,\\
\ $-5.2 (-3.0) [-2.0] \times 10^{-5}$ $< k_0^m/\Lambda^2 <$ $4.6 (2.8) [1.8] \times 10^{-5}$ GeV$^{-2}$,\\
\ $-4.8 (-2.8) [-2.0] \times 10^{-5}$ $< k_2^w/\Lambda^2 <$ $4.6 (2.8) [2.1] \times 10^{-5}$ GeV$^{-2}$},
\end{itemize}

\begin{itemize}
\item{(2) $\Lambda_u=\infty$ TeV:\\
\ $-1.0 (-0.59) [-0.42] \times 10^{-5}$ $< k_2^m/\Lambda^2 <$ $1.0 (0.57) [0.41] \times 10^{-5}$ GeV$^{-2}$,\\
\ $-1.7 (-0.90) [-0.70] \times 10^{-6}$ $< k_0^w/\Lambda^2 <$ $1.6 (0.90) [0.60] \times 10^{-6}$ GeV$^{-2}$,\\
\ $-3.4\ (-1.8)\ [-1.4] \times 10^{-6}$ $\ < k_0^m/\Lambda^2 <$ $3.2\ (1.8)\ [1.2] \times 10^{-6}$ GeV$^{-2}$,\\
\ $-0.50 (-0.30) [-0.21] \times 10^{-5}$ $< k_2^w/\Lambda^2 <$ $0.50 (0.28) [0.20] \times 10^{-5}$ GeV$^{-2}$,}
\end{itemize}

\begin{itemize}
\item{(3) $-3.7 (-2.2) [-1.7] \times 10^{-5}$ GeV$^{-2}$ $< a_n/\Lambda^2 <$ $3.9 (2.4) [2.0] \times 10^{-5}$ GeV$^{-2}$.}
\end{itemize}

\begin{figure}[ht]
\centering
\subfigure[]{
    \label{k2wk2m}
    \includegraphics[width=0.40\textwidth]{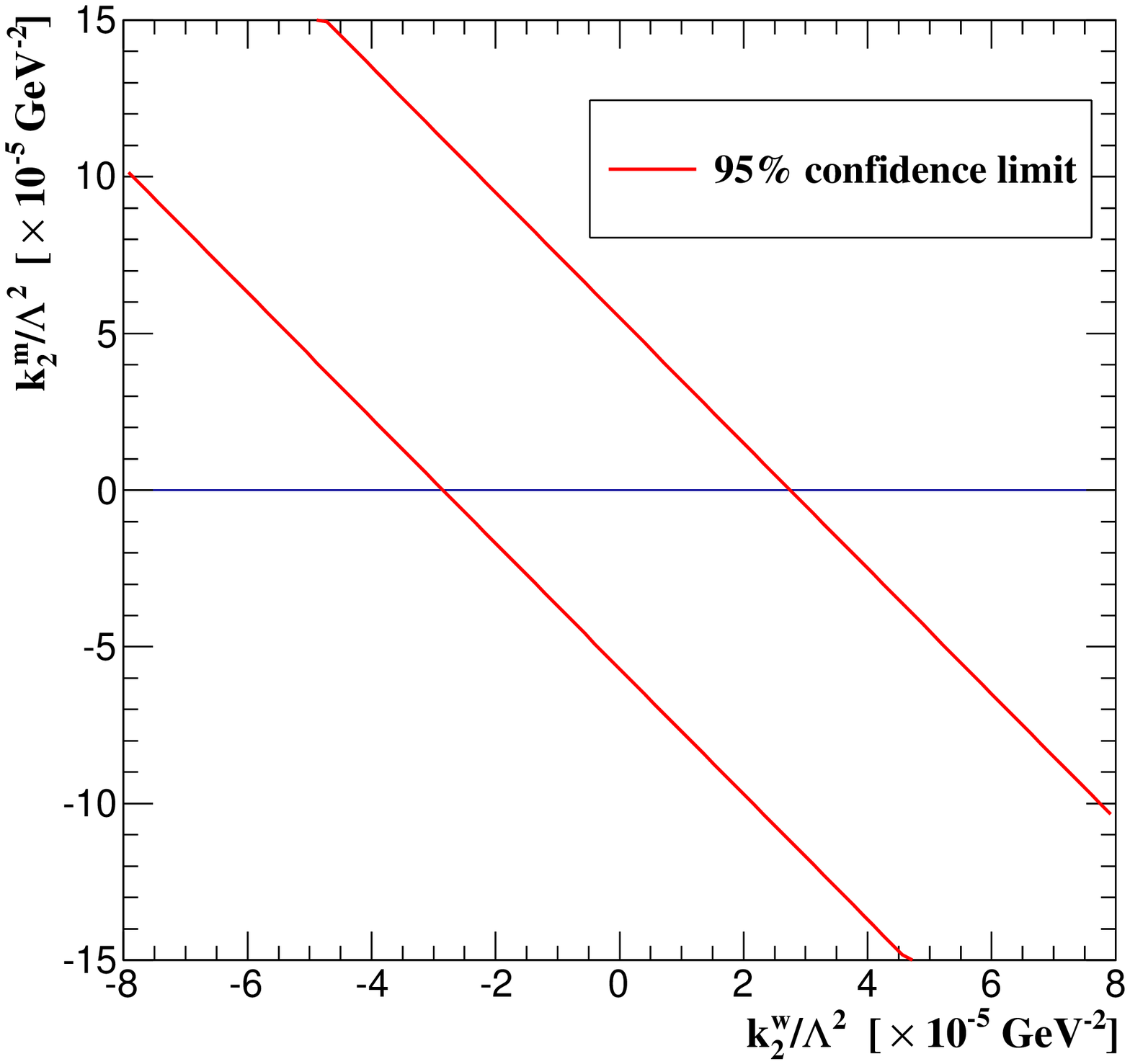}
}
\subfigure[]{
    \label{k0wk2w}
    \includegraphics[width=0.40\textwidth]{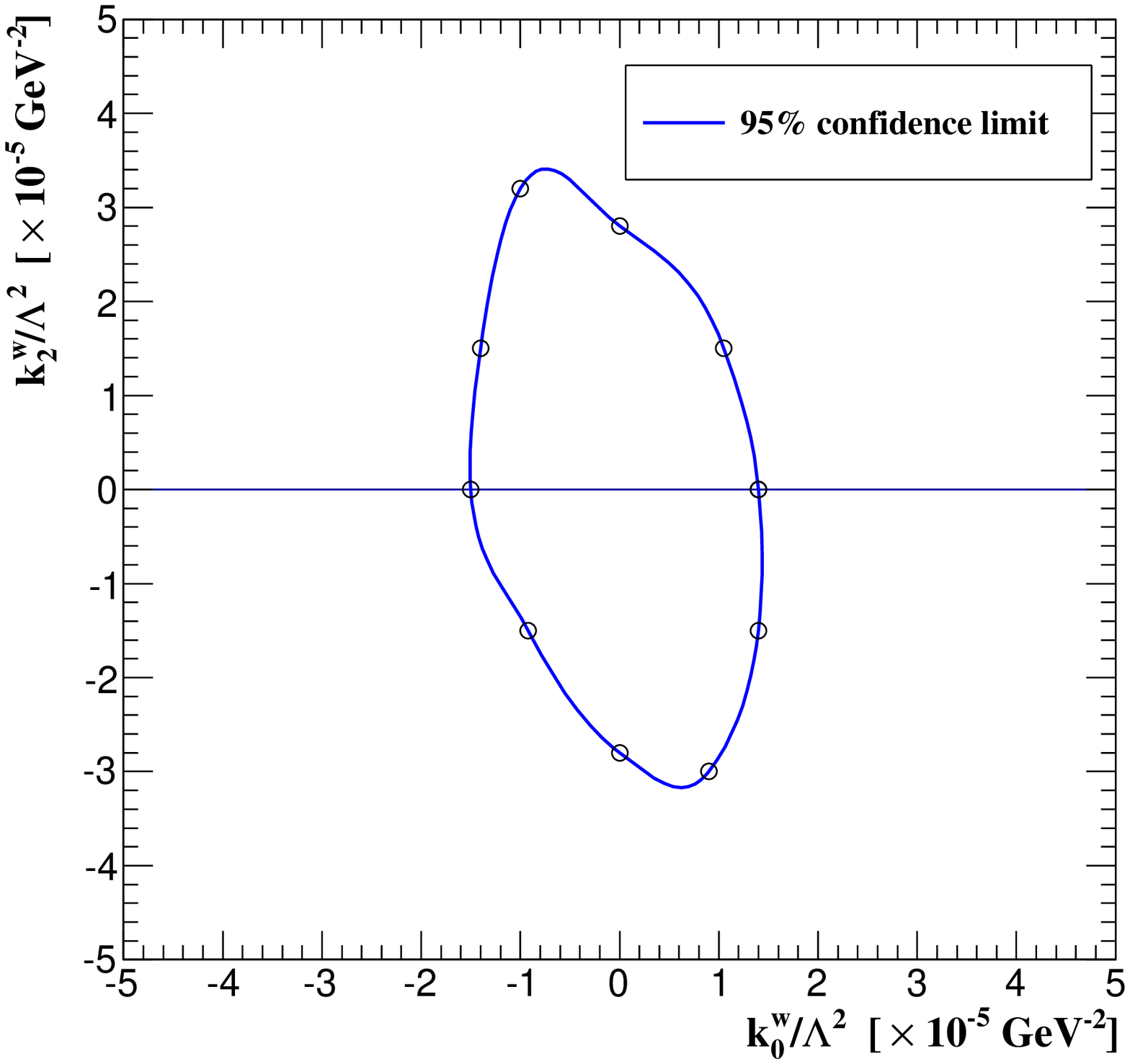}
}
\caption{\label{Correlation} Correlations between anomalous coupling constants: 2$\sigma$ contours for (a) $k^w_2$ and $k^m_2$, and (b) $k^w_0$ and $k^w_2$. Here we assume the ${\cal CP}$ conserving Lagrangian with the form factor of n=5 and $\Lambda_u=2.5$ TeV, as well as the 14TeV LHC with 100 \fbinv\ of data.
}
\end{figure}

Fig.~\ref{Correlation} illustrates the correlation between two coupling constants. The left subplot is for the case of $k^w_2$ and $k^m_2$, as they always appear as the sum $k^w_2+\frac{1}{2}k^m_2$, the 2$\sigma$ contour is simply band-like. The right subplot is for $k^w_0$ and $k^w_2$, where the contour is more complex as a circle . Other correlations can be deduced from these two examples, for more details see Appendix \ref{appendixA}.

Our results can be directly compared with the former experimental results and MC expectations. In Ref.~\cite{Eboli:0310141} and~\cite{Belanger:2000}, bounds on $k^m_2/\Lambda^2$ were derived through the MC simulations via VBF channel at the LHC and $e^+e^- \rightarrow W^+W^-\gamma,Z\gamma\gamma,ZZ\gamma$ processes, respectively. While in Ref.~\cite{Abbiendi:1999aa,Abdallah:2003xn,Acciarri:2000en} (LEP experiments results) and~\cite{Sahin} (MC expectations at $\gamma\gamma$ colliders), limits were given on $a_n/\Lambda^2$. Summary of those results are listed in Table.~\ref{tabK2M} and Table.~\ref{tabAN}. At the 14 TeV LHC, we can set more stringent limit by at least three orders of magnitude compared to LEP results. Although a bit worse than the VBF channel, the leptonic decay mode of \wza has simpler event topology and may be less contaminated by the QCD and VBF systematics.

\begin{center}
\begin{table*}[h!]
\begin{tabular}{|c||c|c|c|}
\hline
\multirow{2}{*}{parameter} 
& \multicolumn{3}{c}{ 95\% confidence interval } \vline\\ 
\cline{2-4}
& {\tiny \wza ($\times 10^{-5}$GeV$^{-2}$)} & {\tiny MC VBF~\cite{Eboli:0310141} ($\times 10^{-5}$GeV$^{-2}$)} & {\tiny MC at LEP2~\cite{Belanger:2000} ($\times 10^{-2}$GeV$^{-2}$)}\\\hline 
$k_2^m/\Lambda^2$              & $[-5.7,5.5]$     &  $[-2.7,2.7]$     &     $[-6.2,6.4]$           \\\hline  
%$k_2^w/\Lambda^2$              & $[-2.8,2.8]$     &  $[-2.7,2.7]$     &     $[-3.1,3.2]$           \\\hline 
%$k_0^w/\Lambda^2$              & $[-1.5,1.4]$     &  $[-0.39,0.39]$     &     $[-0.78,0.73]$           \\\hline  
%$k_0^m/\Lambda^2$              & $[-3.0,2.8]$     &  $[-0.39,0.39]$     &     $[-1.2,1.1]$           \\\hline 
\end{tabular}
\caption{\label{tabK2M} Comparison of 95\% C.L. limits on $k_2^m/\Lambda^2$, between us and previous literatures~\cite{Eboli:0310141,Belanger:2000}, with an integrated luminosity of 100 $\fbinv$ at the 14 TeV LHC. The same form factor has been applied with $n=5$ and $\Lambda_u=2.5$ TeV (Eq.~(\ref{k2mL})).
Note the results from Ref.~\cite{Eboli:0310141,Belanger:2000} assume that all $k^j_i$s (Eq.~(\ref{aqgceff})) are mutually independent, while we take the 4-dimensional parametrization as mentioned before in Sec.~\ref{effwwza} which leads to geniune \wwza aQGC but not e.g. \wwaa ones.}
\end{table*}
\end{center}

\begin{center}
\begin{table*}[h!]
\begin{tabular}{|c||c|c|c|c|c|}
\hline
\multirow{2}{*}{parameter} 
& \multicolumn{5}{c}{ 95\% confidence interval } \vline\\ 
\cline{2-6}
& {\tiny \wza ($\times 10^{-5}$GeV$^{-2}$)} & {\tiny OPAL~\cite{Abbiendi:1999aa}} & {\tiny DELPHI~\cite{Abdallah:2003xn}} & {\tiny L3~\cite{Acciarri:2000en}} & {\tiny MC at $\gamma\gamma$ collider~\cite{Sahin}} \\\hline 
$a_n/\Lambda^2$              & $[-2.2,2.4]$     &  $[-0.61,0.57]$     &     $[-0.18,0.14]$    & $[-0.41.0.37]$     & $[-0.03,0.03]$       \\\hline  
\end{tabular}
\caption{\label{tabAN} Comparison of 95\% C.L. limits on $a_n/\Lambda^2$. The integrated luminosity of \wza corresponds to 100 $\fbinv$ at the 14 TeV LHC. For $\gamma\gamma$ collider simulations~\cite{Sahin}, the value relies on polarization of the beam and $\sqrt{s}$, and we pick the best value.}
\end{table*}
\end{center}

\section{Conclusion and Discussion}
\label{discuss}

The future upgrade of LHC with higher center of mass energy and luminosity enables measurement of triple gauge boson production and anomalous quartic gauge couplings, and \wza production will be a potential channel which can be exploited to test the SM predictions and probe \wwza anomalous coupling exclusively with lower background contamination.

In summary, our study shows that at the 14 TeV LHC with an integrated luminosity of 100 \fbinv, one can reach a significance of about 3$\sigma$ to observe the SM \wza production, and can constrain at 95\% C.L. the anomalous \wwza coupling parameters, e.g.,  $k^m_2/\Lambda^2$ and  $a_n/\Lambda^2$ at $1 \times 10^{-5} \text{GeV}^{\text{-2}}$, respectively. The expected limits are far beyond the existing LEP results, and can be comparable with the ones from VBF MC simulation studies~\cite{Eboli:0310141}.

%--------------------------------------------------------------------
\acknowledgments
This work is supported in part by the National Natural Science Foundation of China, under Grants No. 10721063, No. 10975004, No. 10635030 and No. 11205008, and National Fund for Fostering Talents in Basic Science, under Grant No. J1103206.
%-----------------------------------------------------------------
\appendix
\section{Appendix: a more general parametrization to genuine \wwza aQGCs}
\label{appendixA}
\ \ \ \ Taking account all the coefficients that characterize each anomalous coupling, we have the following set of equations~\cite{Eboli:0310141}:
\begin{equation}
k^{\gamma}_i=k^w_i+k^b_i+k^m_i,\ \ i=0,c,1,
\label{1}
\end{equation}
\begin{equation}
k^{\gamma}_{23}=k^w_2+k^b_2+k^m_2+k^w_3+k^m_3
\label{2}
\end{equation}
\begin{equation}
k^Z_0=\frac{c_w}{s_w}(k^w_0+k^w_1)-\frac{s_w}{c_w}(k^b_0+k^b_1)+c_{zw}(k^m_0+k^m_1)
\label{3}
\end{equation}
\begin{equation}
k^Z_c=\frac{c_w}{s_w}(k^w_c+k^w_2+k^w_3)-\frac{s_w}{c_w}(k^b_c+k^b_2)+c_{zw}(k^m_c+k^m_2+k^m_3)
\label{4}
\end{equation}
\begin{equation}
k^W_0=\frac{c_w}{s_w}k^w_0-\frac{s_w}{c_w}k^b_0+c_{zw}k^m_0
\label{5}
\end{equation}
\begin{equation}
k^W_c=\frac{c_w}{s_w}k^w_c-\frac{s_w}{c_w}k^b_c+c_{zw}k^m_c
\label{6}
\end{equation}
\begin{equation}
k^W_j = k^w_j + \frac{1}{2}k^m_j,\ \ (i=1,2,3).
\label{7}
\end{equation}

Here the first four set of equations, (\ref{1}) - (\ref{4}), are the coupling constants of $WW\gamma\gamma$, $ZZ\gamma\gamma$, $ZZZ\gamma$ and thus are irrelevant to \wwza coupling, and we would like to find a solution for which they vanish. Here we seek for restrictions on these parameters that can lead to zero of Eqs. (\ref{1}) - (\ref{4}), which are:
\begin{equation}
k^b_1=0,\ \ k^b_2=0,
\label{6-1}
\end{equation}
\begin{equation}
k^w_0=k^w_c,\ \ k^m_0=k^m_c,\ \ k^b_0=k^b_c,
\label{6-2}
\end{equation}
\begin{equation}
k^m_1=k^m_2+k^m_3,\ \ k^w_1=k^w_2+k^w_3,
\label{6-3}
\end{equation}
\begin{equation}
k^w_0+k^b_0+k^m_0=0,
\label{6-4}
\end{equation}
\begin{equation}
\frac{c_w}{s_w}(k^w_0+k^w_1)-\frac{s_w}{c_w}(k^b_0+k^b_1)+c_{zw}(k^m_0+k^m_1)=0,
\label{6-5}
\end{equation}
\begin{equation}
k^w_1+k^m_1=0.
\label{6-6}
\end{equation}

The conditions Eqs.(\ref{6-4}-\ref{6-6}), lead to $2k^w_0+k^m_0+k^w_1=0$. We have here in total 10 independent restrictions, hence leaving 4 independent variables. And we choose them to be $k^w_0, k^m_0, k^w_2$, and $k^m_2$. It is then easy to verify that the couplings constants can be expressed exactly as those in Eq. (\ref{newpar1}) - Eq. (\ref{newpar5}).

Having obtained the paramterization of genuine \wwza aQGC, we may further investigate the correlations between two paramters while setting the remaining two zero, which involve a total of six combinations. We take the following two combinations as examples, and the other cases can be inferred in a similar manner:
\begin{itemize}
\item (1) $k^m_2,\ k^w_2 \neq 0$.
\end{itemize}

It is easy to verify that we have $k^W_0=0,k^W_c=0,k^W_1=0$ and $k^W_2=k^w_2+\frac{1}{2}k^m_2,k^W_3=-(k^w_2+\frac{1}{2}k^m_2)$. Observe that this special case is equivalent to Eq. (\ref{k2mL}), if we subsititute $k^m_2$ in (\ref{k2mL}) with $2k^w_2+k^m_2$. Thus restrictions on $k^m_2$ and $k^w_2$ can be expressed as (for 100 \fbinv\ LHC and with ff n=5, for instance):
\begin{equation}
-5.7 \times 10^{-5} < (2k^w_2+k^m_2)/\Lambda^2 < 5.5 \times 10^{-5}{\rm GeV}^{-2}.
\end{equation}
One may as well check that when only $k^m_0$ and $k^w_0$ are left non-zero, the confidence region can be extracted directly if we make substitution $k^w_0 \rightarrow k^w_0+\frac{1}{2}k^m_0$ in the inequalities given in Sec. \ref{wwzaaQGC}.

\begin{itemize}
\item (2) $k^w_0, k^w_2 \neq 0$.
\end{itemize}

In fact we will see this is the only case we need to consider, where each coefficient $k^W_i$ is written as:
\begin{equation}
k^W_0=\frac{1}{c_ws_w}k^w_0,
\label{2par1}
\end{equation}
\begin{equation}
k^W_c=\frac{1}{c_ws_w}k^w_0,
\end{equation}
\begin{equation}
k^W_1=-k^w_0,
\end{equation}
\begin{equation}
k^W_2=k^w_2,
\end{equation}
\begin{equation}
k^W_3=-k^w_0-k^w_2.
\label{2par5}
\end{equation}
Other cases can be related to above with simple substitutions. For example, when $k^m_0, k^m_2 \neq 0$, one just needs to rewrite Eq.(\ref{2par1}) - (\ref{2par5}) with $k^w_0 \rightarrow \frac{1}{2}k^m_0$, $k^w_2 \rightarrow \frac{1}{2}k^m_0$.

%%%%%%%%%%%%%% Begin References %%%%%%%%%%%%%%%%%%%%%%%%%%%%%%%%%%%%%%%%

\end{document}